\documentclass[journal=jctcce,a4paper,manuscript=article]{achemso}

\usepackage{amsfonts}
\usepackage{amsmath}
\usepackage{amssymb}
\usepackage{amsthm}
\usepackage[american]{babel}
\usepackage{braket}
\usepackage{float}
\usepackage{graphicx} 
\usepackage{here}
\usepackage{multirow}
\usepackage{siunitx}
\usepackage{url}
\usepackage{xcolor}
\setcitestyle{square}

\title{Non-iterative Triples for\\ Transcorrelated Coupled Cluster Theory}
\author{Maximilian M\"orchen}
\affiliation{ETH Z{\"u}rich, Department of Chemistry and Applied Biosciences, Vladimir-Prelog-Weg 2, 8093 Z{\"u}rich, Switzerland}
\author{Alberto Baiardi}
\affiliation{ETH Z{\"u}rich, Department of Chemistry and Applied Biosciences, Vladimir-Prelog-Weg 2, 8093 Z{\"u}rich, Switzerland}
\author{Micha\l{} Lesiuk}
\affiliation{University of Warsaw, Faculty of Chemistry, Pasteura 1, 02-093 Warsaw, Poland}
\author{Markus Reiher}
\affiliation{ETH Z{\"u}rich, Department of Chemistry and Applied Biosciences, Vladimir-Prelog-Weg 2, 8093 Z{\"u}rich, Switzerland}

\email{mreiher@ethz.ch}
\date{January 25, 2025}

\newcommand{\cre}[1]{\hat{a}_{#1}^{\dagger}}
\newcommand{\ann}[1]{\hat{a}_{#1}^{}}

\begin{document}
\maketitle

\begin{abstract}
    We present an implementation of a perturbative triples correction for the coupled cluster ansatz including single and double excitations based on the transcorrelated Hamiltonian.
    Transcorrelation introduces explicit electron correlation in the electronic Hamiltonian through similarity transformation with a correlation factor.
    Due to this transformation, the transcorrelated Hamiltonian includes up to three-body couplings and becomes non-Hermitian. 
    Since the conventional coupled cluster equations are solved by projection, it is well suited to harbor non-Hermitian Hamiltonians.
    The arising three-body operator, however, creates a huge memory bottleneck and increases the runtime scaling of the coupled cluster equations. 
    As it has been shown that the three-body operator can be approximated, by expressing the Hamiltonian  in the normal-ordered form, we investigate this approximation for the perturbative triples correction.
    Results are compared with a code-generation based transcorrelated coupled cluster implementation up to quadruple excitations.
\end{abstract}

\section{Introduction}
The accurate description of electron correlation remains one of the greatest challenges in quantum chemistry. 
Within a one-particle picture, the energy contribution of electron correlation can be defined as the difference between the exact full configuration interaction (FCI), and the mean-field Hartree--Fock (HF) energy. 
One of the most powerful approaches to recover the correlation energy is the coupled cluster\cite{Coester_1958, Coester_1960, Cizek_1966, Cizek_1971} (CC) approach. 
In CC, the wave function is parametrized by an exponential cluster operator acting on a reference wave function (usually the HF determinant). 
In practice, the respective wave function expansion is truncated after single and double excitations\cite{Purvis_1982, Scuseria_1987} (CCSD) to avoid the exponential scaling. 
The inclusion of full triple excitations (CCSDT)\cite{Noga_1987, Scuseria_1988} can significantly improve on the accuracy of the correlation energy\cite{Lesiuk_2022}, but accounting for it is computationally too demanding for most system sizes.
Hence, triples are usually approximated by perturbative approaches\cite{Raghavachari_1989} (in particular in the CCSD(T) model) as a good compromise between cost and accuracy. 

The size of the one-particle basis, given by a set of Gaussian functions, determines the overall quality of the a calculation.
With an increasing number of Gaussians, the basis set eventually converges to the complete basis set limit.
However, due to the high computational scaling of most \textit{ab initio} methods, it is not always feasible to apply large basis sets. 

Explicitly correlated methods are a powerful alternative in this context as they enforce the electron-electron cusp condition~\cite{Kato_1957, helgaker_molecular_2000} by explicitly incorporating the electronic distance into the wave function.
As a result, the convergence of the results toward the complete basis set limit is greatly accelerated.
For most applications, the ans\"atze of the R12/F12 family~\cite{Haettig_2012,Kong_2012,Tenno_2012} have become standard tools of providing a computationally tractable way to introduce explicit correlation at the level of electronic pair functions.
However, a significant drawback of these methods is the nontrivial generalization to higher-than-double excitations.
This is an important deficiency as it is well-known that at least triple excitations are required to achieve chemical accuracy for most properties.
In fact, in most implementations of the F12 variant of the popular CCSD(T) method, triply excited contributions to the energy are evaluated using the standard orbital formula, i.e., without any form of explicit correlation.
In other words, only the CCSD-F12 component of the energy benefits from the accelerated convergence to the basis set limit, while the (T) component is not improved in a meaningful way.
Clearly, this is not a satisfactory situation as the basis set error in the (T) contribution will likely dominate in the overall error budget. 

A natural route to solve this problem is to extend the F12 theory to triple (and possibly higher) excitations.
This approach was adopted by K\"ohn\cite{Kohn2009, Koehn_2009_f12Triples, kohn2010} who proposed a modified ansatz for explicitly correlated CC wave functions which incorporates explicitly correlated components into the connected triple excitations using a set of cusp conditions (extended SP ansatz).
It was demonstrated that this method accelerates the convergence of the (T) correction to the complete basis set limit, but at a significantly increased cost in comparison with the standard approach.
A much simpler approach to address the aforementioned problem of lack of explicit correlation in the (T) correction is based on scaling it by the ratio of MP2-F12 and MP2 energies in a given basis (see Refs.~\citenum{Knizia_2009} and \citenum{Adler_2007} and references therein for an extended discussion).
This procedure typically improves the results significantly, but is not free from drawbacks.
First, it breaks size consistency of the results which becomes problematic in applications to large systems.
Second, it assumes that the MP2 and (T) energies converge to the CBS limit at the same rate.
However, MP2 energies tend to converge more slowly, and hence, the scaled (T) correction is typically an overestimation.
These problems were addressed in the recent work of K\'allay and collaborators~\cite{kallay2021}.
They propose to perform the scaling not at the level of the total (T) correction, but divide it into contributions from individual triplets of occupied orbitals ($ijk$) and scale each such contribution separately.
While this approach mitigates the size-consistency issue, it is still based on the assumption that MP2 and (T) energies converge to the CBS limit at the same rate.

A different approach to explicitly correlate electrons is transcorrelation, introduced by Boys and Handy\cite{Boys_Handy_1969_Transcorrelation}, which we adopt in this study.
In contrast to including the interelectronic distance into the wave function, the Hamiltonian is similarity transformed by a correlation factor. 
As a result, transcorrelation is generally applicable, even to multiconfigurational wave functions \cite{Baiardi_2020,Baiardi_2022,Cohen_2019,Liao_2023}.
However, the resulting transcorrelated Hamiltonian is no longer Hermitian and includes a three-body operator. 
Recently, the transcorrelation idea has experienced a vivid revival\cite{Nooijen_1998,Tenno_2000a,Tenno_2000b,Imamura_2003,Umezawa_2003,Umezawa_2004a,Umezawa_2004b,Umezawa_2005,Luo_2010a,Luo_2011,Luo_2012,Luo_2010b,Yanai_2012,Sharma_2014,Ochi_2012,Ochi_2014,Kersten_2016,Cohen_2019,Dobrautz_2019,Luo_2018,Gunter_2021,Ochi_2016,Ochi_2015,Jeszenszki_2018,Liao_2021,Schraivogel_2021,Schraivogel_2023,Luo_2022,Motta_2020,Kumar_2022,Khamoshi_2021,Giner_2021,Dobrautz_2022,Ammar_2022,Liao_2023,Ochi_2023,lee2023a,ammar2023,kats2024,szenes2024}.

In the context of CC, the non-Hermiticity does not impose any issues, since the Hamiltonian is similarity transformed by the exponential cluster operator, resulting already in a non-Hermitian Hamiltonian.
Nevertheless, the full treatment of the three-body operator increases the computational cost and memory scaling. 
Ten-no combined the transcorrelated Hamiltonian with linearized CC\cite{Hino_2002}, where many contractions involving the three-body operator were neglected.
Later Alavi and co-workers applied CCSD to the transcorrelated Hamiltonian (TC-CCSD) for the homogeneous electron gas\cite{Liao_2021}, atoms\cite{Schraivogel_2021} and molecules\cite{Schraivogel_2023}. 
We emphasize, however, that the basis-set limit of TC-CCSD is different from that of CCSD.
In order to deal with the three-body couplings, they introduced an approximation based on the normal-ordered Hamiltonian\cite{Schraivogel_2023}. 
In this approximation, the full TC Hamiltonian is normal-ordered with respect to a single reference determinant and subsequently the remaining three-body integrals are neglected.
Within this procedure, mean-field contributions are transferred from the three-body operator to two-, one-, and zero-body operators.
In the same paper, they introduced a similar approximation based on the $T_1$-dressed Hamiltonian\cite{Koch_1994, DePrince_2013}, which, however, still suffers from the huge memory scaling. 
In the $T_1$-dressed Hamiltonian approximation, all remaining three-body terms are neglected, meaning that effective intermediates are precontracted with the three-body operator before the operator is neglected.
Another approach\cite{Christlmaier_2023} to reduce the drawbacks of the three-body operator is based on the generalized normal-ordering\cite{kutzelnigg1997,kutzelnigg1999} of the three-body operator, which reduces the scaling to $\mathcal{O}(N^5)$. 

For higher-order CC models, the explicit treatment of the three-body operator results in numerous possible diagrams.
A derivation of the CCSD equation for Hamiltonians including three-body couplings was given by Piecuch and co-workers in the context of nuclear structure theory\cite{Hagen_2007}. 
The total number of CCSD diagrams for a Hamiltonian including three-body couplings increases from 48 to 116. 
Hence, tools for automatic code contractions become important for the implementation of TC-CC with higher than doubles excitations. 

In this work, we derive a non-iterative triples approximation for TC-CCSD(T) including three-body operators and the normal-order approximation.
A first CCSD(T) implementation for Hamiltonians including a three-body operator\cite{binder2013} was derived based on the $\Lambda$CCSD(T)\cite{kucharski1998,crawford1998,taube2008a} method.
Later, Kats et al. derived the $\Lambda$CCSD(T) method for the transcorrelated Hamiltonian\cite{kats2024}, however without the inclusion of the full three-body operator.

We implemented TC-CCSDT and TC-CCSDTQ for the full transcorrelated Hamiltonian with the \texttt{Wick\&d}\cite{Evangelista_2022,fevangelista_2023} library to generate the required CC tensor contractions.
In the following section, we first briefly review the conventional non-iterative triples approximation and the transcorrelated Hamiltonian.
Afterward, we show the required diagrams including three-body couplings for perturbative triples.
The correlation factor applied in this work is introduced in Section \ref{Sec:Comp_Meth}.
We compare these TC-CC models for Be, Ne, LiH and Be$_2$ in Section \ref{Sec:Results}.

\section{Theory}
If not stated otherwise, we assume spin-orbital labels for all operators.
For convenience, we summarize our notation in Table \ref{tab:notation}.

\begin{table}[htb]
    \centering
    \begin{tabular}{c c}
        \hline 
        symbol & meaning \\
        \hline\hline
         $i, j, k, l, \dots$ & occupied spin-orbital indices \\
         $a, b, c, d, \dots$ & virtual spin-orbital indices \\
         $p, q, r, s, \dots$ & general spin-orbital indices \\
         $\{\dots\}$         & denotes a normal-ordered string of second quantized operators \\
         $\braket{pq|rs}$    & electron repulsion integrals in physics notation \\
         $\braket{pq||rs}$   & antisymmetrized two-body integrals, e.g. $\braket{pq|rs} - \braket{pq|sr}$ \\
        \hline 
    \end{tabular}
    \caption{Notation in the Current Work.}
    \label{tab:notation}
\end{table}

The non-relativistic normal-ordered Hamiltonian in spin-orbital notation reads
\begin{align}
    \hat{H}_{\text{N}} = \hat{H} - E_{\text{ref}} = 
    \sum_{pq} f^p_q \left\{\cre{p}\ann{q}\right\} + 
    \frac{1}{4} \sum_{pqrs} \braket{pq||rs} \left\{\cre{p}\cre{q}\ann{s}\ann{r} \right\}
\end{align}
where 
\begin{align}
    E_{\text{ref}} = \sum_{i} h^i_i + \frac{1}{2}\sum_{ij} \braket{ij||ij} + E_{N,N}
    \label{eq:eref}
\end{align}
is the mean-field energy contribution of the reference determinant, including the nuclear repulsion energy $E_{N, N}$.
$f^p_q$ is an element of the Fock matrix
\begin{align}
    f^p_q = \left(h^p_q + \sum_{i} \braket{pi||qi}\right) \left\{\cre{p}\ann{q}\right\},
    \label{eq:fockmat}
\end{align}
where $h^p_q$ comprises all one-body operators, i.e. the kinetic energy and electron-nuclear interactions. 
In order to distinguish $\hat{H}_N$ from the transcorrelated Hamiltonian, the former quantity is referred to as the conventional Hamiltonian.
We consistently map the indices of the creation and annihilation operators by placing them in the superscript and subscript, respectively, in the corresponding tensor (bra and ket of antisymmetrized integrals, respectively).

We employ the usual diagrammatic representation\cite{feynman1949,goldstone1957,hugenholtz1957}, where dashed lines represent operators.
For example, two outgoing arcs connected to an ``\textsc{x}'' correspond to a one-body operator, two outgoing arcs connected to another set of two outgoing arcs represent a two-body operator. 
Solid horizontal lines represent amplitudes, where the number of vertices corresponds to the excitation degree of the amplitude.
The dotted line represents the denominator constructed from the diagonal entries of the Fock operator; for example
\begin{align}\label{eq:denom}
    D_{ijk}^{abc} = D^{ijk}_{abc} = f_i^i + f_j^j + f_k^k - f_a^a - f_b^b - f_c^c
\end{align}
is the three-body denominator.
The CC amplitudes are denoted by the usual symbols $t^a_i$, $t_{ij}^{ab}$, $t_{ijk}^{abc}$, and so forth. 
For convenience, we will employ both an algebraic and diagrammatic representation in this work.

\subsection{Conventional Perturbative Triples}
In this section, we rederive the conventional perturbative triples correction to the energy in the formalism of Stanton\cite{Stanton_1997}.
In the subsequent section, this framework is extended to incorporate the three-body operator in the electronic Hamiltonian.

As described by Stanton in Ref. \citenum{Stanton_1997}, the CCSD method with $\hat{T} = \hat{T}_1 + \hat{T}_2$ can be transformed into an eigenvalue problem of $\bar{H}$, where the eigenvector is the reference determinant $\ket{\Phi_0}$
\begin{align}
    e^{-\hat{T}}\hat{H}e^{-\hat{T}} \ket{\Phi_0} = \bar{H} \ket{\Phi_0} = E_{\text{CCSD}} \ket{\Phi_0}
\end{align}
which is valid after projection onto a manifold of singly- and doubly-excited determinants.
However, since the similarity transformation is not unitary, $\bar{H}$ is not Hermitian.
Hence, the left eigenvector is not given by
\begin{align}
    \bra{\Phi_0} \bar{H} \neq \bra{\Phi_0} E_{\text{CCSD}}
\end{align}
In order to evaluate the left eigenvector explicitly, the so-called $\Lambda$ equations have to be solved, so that 
\begin{align}
    \bra{\Phi_0} \mathcal{L} \bar{H} = \bra{\Phi_0} \mathcal{L} E_{\text{CCSD}}
\end{align}
with $\mathcal{L} = 1 + \Lambda$ where $\Lambda$ is a de-excitation operator.
Now, instead of $\hat{H}$, $\bar{H}$ is the zeroth-order Hamiltonian with $\ket{\Phi_0}$ and $\bra{\Phi_0}\mathcal{L}$ as left- and right-hand states, respectively.
The perturbative expansion of the similarity transformed Hamiltonian according to Stanton\cite{Stanton_1997} results in the leading-order correction to the energy
\begin{align}
    \Delta E^{[3]} = 
    \braket{\Phi_0| \mathcal{L} \mathbf{S}}
    \braket{\mathbf{S}| \bar{H}^{[1]} \mathbf{T}} 
    \mathbf{D}_3
    \braket{\mathbf{T}| \bar{H}^{[2]} \Phi_0} +
    \braket{\Phi_0| \mathcal{L} \mathbf{D}}
    \braket{\mathbf{D}| \bar{H}^{[1]} \mathbf{T}} 
    \mathbf{D}_3
    \braket{\mathbf{T}| \bar{H}^{[2]} \Phi_0}
\end{align}
where \textbf{S}, \textbf{D}, and \textbf{T} correspond to singly, doubly, and triply excited determinants, respectively, and \textbf{D\textsubscript{3}} is a diagonal matrix with the inverse of the elements defined in Eq. \eqref{eq:denom} on the diagonal.
The order of the Hamiltonian in the perturbation theory expansion is indicated by the square brackets, i.e. [1] or [2].
This perturbative expansion leads to two different sets of amplitudes which can be evaluated according to  
\begin{align}
    \tilde{t}_{ijk}^{abc} & = \left(D_{ijk}^{abs}\right)^{-1}\braket{\Phi_{ijk}^{abc}| [\hat{H},\hat{T}]\Phi_0}, \\
    \bar{t}_{ijk}^{abc}   & = \left(D_{ijk}^{abs}\right)^{-1}\braket{\Phi_0| \mathcal{L}\hat{H}\Phi_{ijk}^{abc}}.
\end{align}
To reduce the computational cost, $\mathcal{L}$ is usually approximated by $\hat{T}^\dagger$, leading to two intermediates 
\begin{align}
    \tilde{t}_{ijk}^{abc} & = \left(D_{ijk}^{abs}\right)^{-1}\braket{\Phi_{ijk}^{abc}| [\hat{H},\hat{T}]\Phi_0} \label{eq:T_3}, \\
    \bar{t}_{ijk}^{abc}   & = \left(D_{ijk}^{abs}\right)^{-1}\braket{\Phi_0| \hat{T}^\dagger\hat{H}\Phi_{ijk}^{abc}} \label{eq:L_3}. 
\end{align}
Note that due to the commutator in Eq. \eqref{eq:T_3} only connected diagrams can be created, while in Eq. \eqref{eq:L_3} also disconnected diagrams appear.
The tensor contractions for the first and second intermediate have the following form
\begin{align}
    D_{ijk}^{abc}\tilde{t}_{ijk}^{abc} & = 
        \frac{1}{4}\sum_{d}\braket{ab || di} t_{jk}^{cd} + 
        \frac{1}{4}\sum_{l}\braket{al || ij} t_{kl}^{bc} \\
    & = 
    \begin{gathered}\includegraphics[]{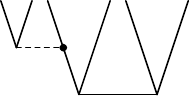}\end{gathered}
    \label{eq:lower_t_3}
\end{align}
and 
\begin{align}
    D^{ijk}_{abc}\bar{t}^{ijk}_{abc} & = 
        \frac{1}{4}\sum_{d}\braket{ab || di} t_{jk}^{cd} + 
        \frac{1}{4}\sum_{l}\braket{al || ij} t_{kl}^{bc} +
        \frac{1}{4}\braket{ab || ij} t_{k}^{c} + 
        \frac{1}{4} f^{a}_{i} t_{jk}^{bc} \\
    & = 
    \begin{gathered}\includegraphics[]{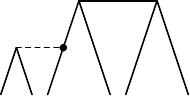}\end{gathered} + 
    \begin{gathered}\includegraphics[]{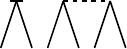}\end{gathered} + 
    \begin{gathered}\includegraphics[]{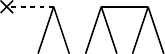}\end{gathered} 
    \label{eq:pert_trip} \\
    & = 
    \begin{gathered}\includegraphics[]{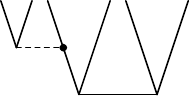}\end{gathered} + 
    \begin{gathered}\includegraphics[]{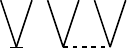}\end{gathered} + 
    \begin{gathered}\includegraphics[]{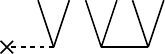}\end{gathered}
    \label{eq:pert_trip_flipped}
\end{align}
respectively.
Note that the last term in the latter equation vanishes due to the Brillouin condition in the case of conventional Hamiltonian, but this is no longer true for the transcorrelated Hamiltonian considered in the subsequent sections.
While the transition from Eq. \eqref{eq:pert_trip} to Eq. \eqref{eq:pert_trip_flipped} is in general not valid, we emphasize here that, due to the antisymmetry of the two-body integrals and the approximation of $\mathcal{L}=T^\dagger$, Eqs. \eqref{eq:pert_trip} and \eqref{eq:pert_trip_flipped} are equivalent.
Therefore, in practice, the latter two diagrams in Eq. \eqref{eq:pert_trip} are added to Eq. \eqref{eq:lower_t_3}, so that the first tensor contraction must be evaluated only once. 
The energy contribution of the non-iterative triples correction for the conventional Hamiltonian can be written as 
\begin{align}
    E_{\text{(T)}} & = \frac{1}{36}\sum_{ijkabc} \bar{t}_{abc}^{ijk} D_{ijk}^{abc} \tilde{t}_{ijk}^{abc} \label{eq:standard_pert_trip}\\
    & = 
    \begin{gathered}\includegraphics[]{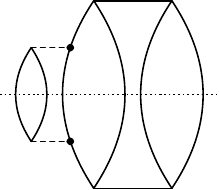}\end{gathered} + 
    \begin{gathered}\includegraphics[]{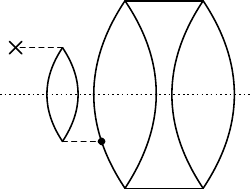}\end{gathered} + 
    \begin{gathered}\includegraphics[]{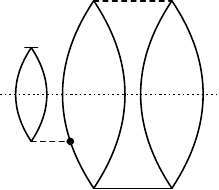}\end{gathered} 
    \label{eq:skeleton}
\end{align}
The skeleton diagrams in Eq. \eqref{eq:skeleton} can be derived by contracting the intermediate amplitudes, $\tilde{t}_{ijk}^{abc}$ and $\bar{t}_{abc}^{ijk}$, with the denominator $D_{ijk}^{abc}$.

\subsection{Transcorrelated Hamiltonian}
The TC Hamiltonian can be derived by similarity transformation of the conventional Hamiltonian with a Jastrow factor\cite{jastrow1955} of the form $e^{f(r_{12})}$ 
\begin{align}\begin{split}
    \hat{H}_{\text{TC}} & =  e^{-f(r_{12})} \hat{H} e^{f(r_{12})} \\
    & = \sum_{pq} h^p_q \cre{p}\ann{q} + 
        \frac{1}{2} \sum_{pqrs} \left(\braket{pq|rs} - K^{pq}_{rs} \right) \cre{p}\cre{q}\ann{s}\ann{r} 
      - \frac{1}{6}\sum_{pqrstu} L_{stu}^{pqr} \cre{p}\cre{q}\cre{r}\ann{u}\ann{t}\ann{s}
\end{split}\end{align}
resulting in the additional two-body term $K^{pq}_{rs}$ and the three-body term $L_{stu}^{pqr}$. 
For the sake of completeness, we summarize some of the important aspects of the TC Hamiltonian.
In the following, we redefine
\begin{align}
    \braket{pq\bar{||}rs} \equiv \braket{pq|rs} - K^{pq}_{rs} - \left(\braket{pq|sr} - K^{pq}_{sr} \right)
\end{align}
Since the two-body term $K_{rs}^{pq}$ is not Hermitian, the redefined, antisymmetrized two-body integrals, lack the symmetry
\begin{align}\label{eq:missing_hertm_2_body}
    \braket{pq\bar{||}rs} \neq \braket{rs\bar{||}pq} 
\end{align}
The antisymmetrized form of the three-body tensor can be defined as
\begin{align}
    \braket{pqr||stu} = 
    L_{stu}^{pqr} -
    L_{sut}^{pqr} -
    L_{tsu}^{pqr} +
    L_{ust}^{pqr} +
    L_{tus}^{pqr} - 
    L_{uts}^{pqr},
\end{align}
such that 
\begin{align}
  \frac{1}{6}\sum_{pqrstu}L^{pqr}_{stu} \cre{p} \cre{q}\cre{r}\ann{u}\ann{t}\ann{s} 
    = \frac{1}{36} \sum_{pqrstu} \braket{pqr||stu} \cre{p}\cre{q}\cre{r}\ann{u}\ann{t}\ann{s} 
\end{align}
After normal-ordering of the three-body term
\begin{align}\begin{split}
    \frac{1}{36} & \sum_{pqrstu}  \braket{pqr||stu}   \cre{p}\cre{q}\cre{r}\ann{u}\ann{t}\ann{s} 
      = \frac{1}{36}\sum_{pqrstu} \braket{pqr||stu} \{\cre{p}\cre{q}\cre{r}\ann{u}\ann{t}\ann{s}\}  \\
    & + \frac{1}{4} \sum_{pqrsi}  \braket{pqi||rsi} \{\cre{p}\cre{q}\ann{s}\ann{r}\}  
      + \frac{1}{2} \sum_{pqij}   \braket{pij||qij} \{\cre{p}\ann{q}\}
      + \frac{1}{6} \sum_{ijk}    \braket{ijk||ijk}
\end{split}\end{align}
the one- and two-body components can be subtracted from the corresponding one- and two-body operators.
Hence, the normal-ordered TC Hamiltonian in the spin-orbital basis reads
\begin{align}\begin{split}
    \hat{H}_{\text{TC, N}} = \hat{H}_{\text{TC}} - \tilde{E}_{\text{ref}} & = 
    \sum_{pq} \tilde{f}_p^q \left\{\cre{p}\ann{q}\right\} + 
    \frac{1}{4} \sum_{pqrs} \braket{pq\tilde{||}rs} \left\{\cre{p}\cre{q}\ann{s}\ann{r} \right\} \\
    & - \frac{1}{36} \sum_{pqrstu} \braket{pqr||stu} \left\{\cre{p}\cre{q}\cre{r}\ann{u}\ann{t}\ann{s} \right\}
\end{split}\end{align}
where 
\begin{align}
    \tilde{E}_{\text{ref}} = \bar{E}_{\text{ref}} - \frac{1}{6} \sum_{ijk} \braket{ijk||ijk}
\end{align}
with
\begin{align}
    \tilde{f}_p^q = \bar{f}_q^p - \frac{1}{2} \sum_{pqij} \braket{pij||qij}
\end{align}
and 
\begin{align}
    \braket{pq \tilde{||} rs} = \braket{pq\bar{||}rs} - \frac{1}{4} \sum_{pqrsi} \braket{pqi||rsi}
\end{align}
include a mean-field contribution from the three-body operator.
Note that $\bar{E}_\text{ref}$ and $\bar{f}^p_q$ can be evaluated through Eq. \eqref{eq:eref} and Eq. \eqref{eq:fockmat}, respectively, by substitution of the conventional antisymmetrized two-body integrals with the corresponding transcorrelated integrals.

To derive the energy and amplitude equations, $\hat{H}_{\text{TC, N}}$ is similarity transformed with the cluster operator.
Because of the new three-body operator, the resulting Baker-Campbell-Hausdorff-series truncates only after the 6-fold commutator. \\
To tackle the memory requirements associated with the three-body operator, the transcorrelated Hamiltonian can be approximated by the normal-ordering approximation\cite{Schraivogel_2021}
\begin{align}\begin{split}
    \hat{H}_{\text{TC, approx.}} = 
    \sum_{pq} \tilde{f}_p^q \left\{\cre{p}\ann{q}\right\} + 
    \frac{1}{4} \sum_{pqrs} \braket{pq\tilde{||}rs} \left\{\cre{p}\cre{q}\ann{s}\ann{r} \right\} 
\end{split}\end{align}

\subsection{Transcorrelated Perturbative Triples}
By application of Eqs. \eqref{eq:T_3} and \eqref{eq:L_3}, which are both linear in the cluster operator (and hence, for example, no $T^2_1$ term appears), we can derive a new set of diagrams for the intermediates of the perturbative triples with the three-body operator.
The following new terms should be added to the standard (T) energy correction formula (with modified integrals):
\begin{align}
    \begin{split}
    D_{ijk}^{abc}\;{}^{(3)}t_{ijk}^{abc} & =
          \frac{1}{36}         \braket{abc||ijk} 
        + \frac{1}{12}\sum_{d }\braket{abc||dij} t_{k}^{d} 
        - \frac{1}{12}\sum_{l }\braket{abl||ijk} t_{l}^{c} \\
      & + \frac{1}{24}\sum_{de}\braket{abc||dei} t_{jk}^{de}
        + \frac{1}{ 4}\sum_{dl}\braket{abl||dij} t_{kl}^{cd}
        + \frac{1}{24}\sum_{lm}\braket{alm||ijk} t_{lm}^{bc} 
    \end{split} \label{eq:add_t3}\\
    & = 
    \begin{gathered}\includegraphics[]{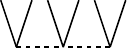}\end{gathered} + 
    \begin{gathered}\includegraphics[]{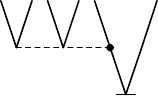}\end{gathered} + 
    \begin{gathered}\includegraphics[]{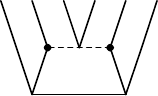}\end{gathered}
\end{align}
Since the contribution of the three-body operator does not result in any disconnected diagrams (such disconnected diagrams would be of an order higher than three) and since the three-body operator is Hermitian, this contribution is identical for both sets of amplitudes ($\bar{t}_{abc}^{ijk}$ and $\tilde{t}^{abc}_{ijk}$).
We note that in the normal-ordering approximation, the additional contributions from Eq. \eqref{eq:add_t3} will vanish.
Adding the new set of diagrams to the intermediates in Eqs. \eqref{eq:pert_trip} and \eqref{eq:lower_t_3} yields the new set of equations:
\begin{align}
    \;{}^{(3)}\tilde{t}_{ijk}^{abc} & = \;{}^{(3)}t_{ijk}^{abc} + \tilde{t}_{ijk}^{abc} \label{eq:inter_1}\\
    \;{}^{(3)}\bar{t}^{ijk}_{abc}   & = \left(\;{}^{(3)}t_{ijk}^{abc}\right)^\dagger + \bar{t}^{ijk}_{abc} \\
    & = \;{}^{(3)}t^{ijk}_{abc} + \bar{t}^{ijk}_{abc} \label{eq:inter_2}
\end{align}
After application of Eqs. \eqref{eq:inter_1} and \eqref{eq:inter_2}, the energy can be evaluated by Eq. \eqref{eq:standard_pert_trip}. 
However, the lack of Hermiticity of the two-body operator in Eq. \eqref{eq:missing_hertm_2_body} has some implications for the formulas. 

First, due to the normal-ordering of the Hamiltonian, the one-body operator is no longer symmetric 
\begin{align}
    \tilde{f}^p_q \neq \tilde{f}^q_p, 
\end{align}
and second, the contractions in Eqs. \eqref{eq:pert_trip} and \eqref{eq:pert_trip_flipped} are not equivalent anymore.
Note that in standard implementations of the (T) correction with the conventional Hamiltonian, only one set of amplitudes, namely $\tilde{t}^{abc}_{ijk}$, is evaluated explicitly. The additional terms present in $\bar{t}^{abc}_{ijk}$ are inexpensive and can be evaluated on the fly, such that $\bar{t}^{abc}_{ijk}$ is never explicitly formed. By contrast, in the implementation of the (T) correction in the transcorrelated framework, both sets of amplitudes ($\bar{t}_{abc}^{ijk}$ and $\tilde{t}^{abc}_{ijk}$) must be separately evaluated. However, we do not expect this to cause a serious memory bottleneck, because these amplitudes can be evaluated in batches with some indices fixed, similarly as in the conventional case.

\section{Computational Methodology} \label{Sec:Comp_Meth}
As the CC method proposed in this work is generally applicable for any correlator, we chose the same correlation factor as in Ref.~\citenum{Baiardi_2022}, namely
\begin{align}
    f(r_{12}) = \frac{1}{2} r_{12} e^{-\gamma r_{12}}
\end{align}
and set $\gamma = 1$.

For the orbital optimization, we followed the strategy described in Ref. \citenum{Baiardi_2022}.
To circumvent the challenges arising from the non-Hermiticity of the transcorrelated Hamiltonian, the orbitals were first optimized by a self-consistent-field procedure for the Hermitian conventional Hamiltonian.
Afterward, the additional two- and three-body terms were added to the Hamiltonian.
We would like to point out that different strategies exist to optimize the orbitals directly for the transcorrelated Hamiltonian (see for example Refs. \citenum{lee2023a} and \citenum{kats2024}).
    
For the calculations of LiH, Be and Be$_2$, we selected the cc-pVDZ, cc-pCVDZ, cc-pVTZ and cc-pVQZ atomic orbital basis sets\cite{Dunning_1989, Prascher_2011} and the cc-pCV\{D, T\}Z\cite{feller1996a, schuchardt2007a} basis from basis set exchange\cite{feller1996a, schuchardt2007a, pritchard2019a}.
We applied the matching cc-pV\{D, T, Q\}Z-RIFIT\cite{Weigend_2002, Haettig_2005} and cc-pCV\{D, T\}Z-RIFIT\cite{kritikou2015a} basis for the density fitting in the integral evaluation.
Because the evaluation of the three-body integrals relies on the resolution of identity, it is named `RI` in the original publication\cite{Baiardi_2022}. 
For the calculations on the Ne-atom, we used the aug-cc-pV\{D, T\}Z\cite{Dunning_1989, kendall1992a} and aug-cc-pwCVDZ\cite{Dunning_1989, kendall1992a, peterson2002a} basis with the corresponding RIFIT basis set for the density fitting in the integral evaluation.
If not stated otherwise, we employ the cc-pV5Z-RIFIT basis set for Li \cite{Haettig_2005} and the uncontracted cc-pV5Z basis for H, Be and Ne \cite{Dunning_1989, Prascher_2011} for the resolution of identity approximation in the three-body integral evaluation.

To implement the amplitude equations in spin-orbital and spin-factorized form, we employ the open-source library \texttt{Wick\&d} \cite{Evangelista_2022, fevangelista_2023}.
Because of the flexibility of this library, it was possible to generate CC equations for any excitation degree and Hamiltonian with $n$-body operators.
Since the underlying algorithm is based on an antisymmetrized form of the tensors with a spin-orbital representation of the creation and annihilation operators, the stored Hamiltonian and amplitudes include many entries, which are zero due to spin.
Especially in the context of higher-order amplitudes or Hamiltonians including a three-body operator, deriving the pure spin-orbital form of the amplitude equations is unfeasible because of the resulting memory and runtime requirements of the generated program.
Hence, we generated the spin-factorized form of the amplitude equations, by creating two fermionic spaces, i.e. $\alpha$- and $\beta$-spin and declaring the required operators accordingly, as described in Ref. \citenum{Evangelista_2022}. 
If TC-CC is approximated by normal-ordering (TC-CC-n), the Hamiltonian includes up to two-body couplings and the diagrams become equal to conventional CC.

The number of generated tensor contractions per CC truncation level can be seen for both spin-orbital and spin-factorized forms in Table
\ref{tab:contr-spatial}.
Due to the large number of required contractions for spin-factorized conventional CCSDT and CCSDTQ, the number of diagrams is enormous for these methods if three-body operators are included. 
Hence, we employed the spin-factorized form for all methods, but TC-CCSDT and TC-CCSDTQ. 

\begin{table}[htb]
    \begin{tabular}{c | r r r r r | r r r r r}
        \hline\hline
                      & \multicolumn{5}{c|}{diagrams (conv.)} & \multicolumn{5}{c}{diagrams (TC)} \\
        CC truncation & E & $T_1$ & $T_2$ & $T_3$ & $T_4$ & E & $T_1$ & $T_2$ & $T_3$ & $T_4$ \\
        \hline
        \multicolumn{11}{c}{Spin-Orbital Basis} \\
        \hline
        2             & 3 &    14 &    31 &       &       & 5 &    29 &    82 &       &       \\
        3             & 3 &    15 &    37 &    47 &       & 6 &    35 &   111 &   224 &       \\
        4             & 3 &    15 &    38 &    53 &   74  & 6 &    36 &   117 &   253 &   410 \\
        \hline
        \multicolumn{11}{c}{Spin-Factorized Basis} \\
        \hline
        2             & 8 & 52 & 172 &     &      & 18 & 164 & 740 &    &    \\
        3             & 8 & 58 & 222 & 520 &      &    &     &     &    &    \\
        4             & 8 & 58 & 231 & 592 & 1289 &    &     &     &    &    \\
        \hline\hline
    \end{tabular}
    \caption{Number of Energy and Amplitude Diagrams for Conventional ('conv.') CC and Transcorrelated ('TC') CC in a Spin-Orbital and Spin-Factorized Basis.}
    \label{tab:contr-spatial}
\end{table}

\section{Results}\label{Sec:Results}
We first investigated the basis set convergence of CCSD, CCSD(T), CCSDT and CCSDTQ with the conventional and transcorrelated Hamiltonian for the Be atom. 
Second, we compared the difference between canonical transcorrelated F12 (CT-F12)\cite{Yanai_2012, Masteran2023} and transcorrelated energy contributions for the Ne atom. 
Then, we examined the dissociation of the LiH molecule in different basis sets with CCSD, CCSD(T), CCSDT, and CCSDTQ based on the conventional and transcorrelated Hamiltonians.
Finally, we studied the Be$_2$ dimer with CCSD, CCSD(T) and CCSDT, as Be$_2$ is a prototypical system that requires at least perturbative triples and a triple-$\zeta$ basis set to be described correctly. 
All CC energies were converged up to $10^{-7}$ hartree.

\subsection{Atomic System: Be}
Since the Be atom is a four electron system, CCSDTQ yields FCI accuracy. 
We compared the basis set convergence with 
a highly accurate result from Ref. \citenum{Puchaslski_2013}.
The normal-ordering approximation of the three-body and the effect of the CC truncation for different basis sets are shown in Table \ref{tab:Be_abs}.

\begin{table}[htb]
    \centering
    {\small
    \begin{tabular}{l l | c c c c c c}
        \hline \hline
        basis    & Hamiltonian & HF         & TC-MF      & SD         & (T)        & T          & Q           \\
        \hline
        cc-pVDZ  & conv.       & -14.572481 &          - & -14.617532 & -14.617570 & -14.617570 & -14.617572  \\
        cc-pVDZ  & TC-n        & -14.572481 & -14.624054 & -14.656796 & -14.656787 & -14.656803 & -14.656806  \\
        cc-pVDZ  & TC          & -14.572481 & -14.624054 & -14.656813 & -14.656805 & -14.656821 & -14.656823  \\
        cc-pVTZ  & conv.       & -14.572876 &          - & -14.623581 & -14.623812 & -14.623823 & -14.623832  \\
        cc-pVTZ  & TC-n        & -14.572876 & -14.624412 & -14.658693 & -14.658747 & -14.658791 & -14.658797  \\
        cc-pVTZ  & TC          & -14.572876 & -14.624412 & -14.658713 & -14.658767 & -14.658809 & -14.658816  \\
        cc-pCVDZ & conv.       & -14.572340 &          - & -14.651582 & -14.651822 & -14.651830 & -14.651816  \\
        cc-pCVDZ & TC-n        & -14.572340 & -14.623911 & -14.670183 & -14.670115 & -14.670155 & -14.670159  \\
        cc-pCVDZ & TC          & -14.572340 & -14.623911 & -14.670206 & -14.670138 & -14.670175 & -14.670179  \\
        cc-pCVTZ & conv.       & -14.572874 &          - & -14.661995 & -14.662435 & -14.662456 & -14.662462  \\
        cc-pCVTZ & TC-n        & -14.572874 & -14.624410 & -14.667613 & -14.667612 & -14.667703 & -14.667710  \\
        cc-pVQZ  & conv.       & -14.572973 &          - & -14.639631 & -14.640128 & -14.640161 & -14.640169  \\
        \hline
        CBS FCI\cite{Puchaslski_2013} & \multicolumn{7}{c}{-14.667356} \\
        \hline \hline
    \end{tabular}
    }
    \caption{Ground-State Energy in hartree of the Be Atom for Different Basis Sets and Methods. 
    The second column denotes the Hamiltonian: conventional ('conv.') or transcorrelated ('TC'). 
    'TC-MF' refers to the mean-field contribution of the transcorrelated Hamiltonian.
    The '-n' suffix indicates the normal-ordering approximation and 'CBS FCI' refers to the reference energy from Ref. \citenum{Puchaslski_2013}.}
    \label{tab:Be_abs}
\end{table}

\begin{table}[htb]
    \centering
    \begin{tabular}{l l | c c c c c c}
        \hline \hline
        Basis    & Hamiltonian & HF         & TC-MF      & SD         & (T)        & T          & Q           \\
        \hline
        cc-pVDZ  & conv.       & -14.572481 &          - &  -0.045051 &  -0.000038 &  -0.000038 &  -0.000002  \\
        cc-pVDZ  & TC-n        & -14.572481 &  -0.051573 &  -0.032742 &   0.000009 &  -0.000008 &  -0.000003  \\
        cc-pVDZ  & TC          & -14.572481 &  -0.051573 &  -0.032759 &   0.000008 &  -0.000008 &  -0.000003  \\
        cc-pVTZ  & conv.       & -14.572876 &          - &  -0.050705 &  -0.000231 &  -0.000242 &  -0.000009  \\
        cc-pVTZ  & TC-n        & -14.572876 &  -0.051536 &  -0.034281 &  -0.000054 &  -0.000098 &  -0.000006  \\
        cc-pVTZ  & TC          & -14.572876 &  -0.051536 &  -0.034301 &  -0.000054 &  -0.000096 &  -0.000006  \\
        cc-pCVDZ & conv.       & -14.572340 &          - &  -0.079242 &  -0.000241 &  -0.000248 &   0.000014  \\
        cc-pCVDZ & TC-n        & -14.572340 &  -0.051571 &  -0.046273 &   0.000068 &   0.000028 &  -0.000004  \\
        cc-pCVDZ & TC          & -14.572340 &  -0.051571 &  -0.046296 &   0.000068 &   0.000031 &  -0.000004  \\
        cc-pCVTZ & conv.       & -14.572874 &          - &  -0.089120 &  -0.000440 &  -0.000461 &  -0.000006  \\
        cc-pCVTZ & TC-n        & -14.572874 &  -0.051535 &  -0.043203 &   0.000001 &  -0.000090 &  -0.000007  \\
        cc-pVQZ  & conv.       & -14.572973 &          - &  -0.066658 &  -0.000497 &  -0.000529 &  -0.000008  \\
        \hline \hline
    \end{tabular}
    \caption{Ground-State Energy Contributions in hartree of the Be Atom for Different Basis Sets and Methods. 
    The second column denotes chosen Hamiltonian: conventional ('conv.') or transcorrelated ('TC'). 
    The '-n' suffix indicates the normal-ordering approximation.
    'TC-MF' refers to the additional mean-field contribution of the transcorrelated Hamiltonian.}
    \label{tab:Be_contr}
\end{table}

Table \ref{tab:Be_abs} shows the expected behavior of the total electronic energy for the valence basis sets with respect of
CC truncation level and basis set size: the inclusion of perturbative triples improved the CCSD energies, with a further improvement by the full triples. 
The impact of the perturbative triples increased with basis set size, reducing the error of CCSD with respect to CCSDTQ from 0.04 to 0.002 millihartree, 0.251 to 0.02 millihartree, and 0.538 to 0.041 millihartree, for the double-, triple-, and quadruple-$\zeta$ basis, respectively. 
In contrast to CCSD(T), CCSDT only yielded a negligibly small further improvement.
For the transcorrelated Hamiltonian in double-$\zeta$ basis, the application of the perturbative triples slightly detoriated the CCSD energies with respect to CCSDTQ. 
However, the absolute error for CCSD is only 0.01 millihartree and for CCSD(T) 0.019 millihartree with respect to CCSDTQ. 
In the triple-$\zeta$ basis, CCSD(T) improves on the CCSD energies, reducing the error from 0.103 to 0.048 millihartree and 104 to 0.05 millihartree for the full transcorrelated and normal-ordering approximated methods, respectively. 
Inspecting the absolute energies, it becomes also evident, that transcorrelated energies
can approach the CBS limit from below; the results for TC-CCSDTQ in the core-valence double-$\zeta$ basis turned out to be 2.822 millihartree below the CBS result. 
However, the error of TC-CCSDTQ-n was found to be only 0.354 millihartree for the larger core-valence triple-$\zeta$ basis because the correlation energy contribution due to TC-CCSD decreases in contrast to the corresponding double-$\zeta$ basis results, which could suggest counterbalancing because of transcorrelation. 

Table \ref{tab:Be_contr} shows (analogously to Ref. \citenum{Werner2010}) the pair correlation (TC-MF), which is already present at the mean-field level.
Note that the pair correlation captures the largest part of the correlation energy and changes only slightly with increasing basis set size.
Also the inclusion of core orbitals has a small effect on this contribution.
In contrast to the conventional Hamiltonian, the contribution from TC-CCSD is lower, because some correlation effects are already captured by transcorrelation. 
This can also be seen in the contribution of perturbative triples and full triples, as it is significantly lower for the transcorrelated Hamiltonian.
As expected, however, the contributions from the different CC truncation levels increases with increasing basis set size and especially with the inclusion of core orbitals.
The normal-ordering approximation yielded results similar to those obtained with the full transcorrelated Hamiltonian, independent of the CC truncation level, while reducing the overall memory scaling.

\subsection{Atomic System: Ne}
In this section, we compare results obtained with the transcorrelated Hamiltonian with the CT-F12 Hamiltonian with projected Slater-type geminals by Yanai and Shiozaki\cite{Yanai_2012}. 
We refer the reader to the original paper\cite{Yanai_2012} and to Ref. \citenum{Masteran2023} for a detailed description of the CT-F12 Hamiltonian.
  \begin{table}[h!]
      \centering
      \begin{tabular}{l|c c c}
        \hline
         Method                          & aug-cc-pVDZ & aug-cc-pwCVDZ & aug-cc-pVTZ \\
         \hline \hline                                   
         TC-MF                           &   -0.705934 &     -0.706070 &   -0.703358 \\ 
         CT-F12-MF\cite{Masteran2023}    &   -0.111555 &             - &   -0.042846 \\ 
         TC-CCSD-n                       &   -0.049641 &      0.160387 &    0.167318 \\ 
         CT-F12-CCSD\cite{Masteran2023}  &   -0.195575 &             - &   -0.267544 \\ 
         TC-CCSD(T)-n                    &   -0.002815 &     -0.002815 &   -0.002805 \\
         TC-CCSDT-n                      &   -0.000198 &      0.001104 &    0.001924 \\
         CT-F12-CCSDT\cite{Masteran2023} &   -0.002606 &             - &   -0.005110 \\ 
         \hline \hline                                   
      \end{tabular}
      \caption{Correlation Energy Contribution of Different Methods for the Ne Atom in hartree. 
      `TC-MF` and `CT-F12-MF` denote the mean-field contribution of the transcorrelated and CT-F12 Hamiltonians, respectively. 
      Note that for the CT-F12 Hamiltonian the two-body operator was explicitly symmetrized to obey the eight-fold permutation symmetry and the orbitals were optimized by iterative diagonalization of the CT-F12 Fock-operator\cite{Yanai_2012} and therefore differ from the orbitals in this study. 
      To compare the CCSD contribution between the transcorrelated and CT-F12 Hamiltonians, we subtracted the CT-F12-MF contribution from the CT-F12-CCSD contribution. 
      We evaluated the CT-F12-CCSDT contribution by subtracting the total energies of CT-F12-CCSDT and CT-F12-CCSD.}
      \label{tab:ne_contr}
  \end{table}
  \begin{table}[h!]
      \centering
      \begin{tabular}{l|c c c c}
         \hline \hline                                   
         Method                          & aug-cc-pVDZ & aug-cc-pwCVDZ & aug-cc-pVTZ & CBS       \\
         \hline                                   
         TC-CCSD-n                       & -129.253376 & -129.041822   & -129.069349 & -128.8631 \\ 
         CT-F12-CCSD\cite{Masteran2023}  & -128.803480 &             - & -128.843663 & -128.8631 \\ 
         TC-CCSD(T)-n                    & -129.256191 & -129.043871   & -129.073134 &        -  \\
         TC-CCSDT-n                      & -129.253574 & -129.040718   & -129.067424 & -128.8694 \\ 
         CT-F12-CCSDT\cite{Masteran2023} & -128.806086 &             - & -128.848773 & -128.8694 \\ 
         \hline \hline                                   
      \end{tabular}
      \caption{Absolute Energies of the Ne Atom for Different Methods in hartree. Note that for the CT-F12 methods, CT-F12-CC includes the CT-F12 contribution. The CBS energies were evaluated, in analogy to Ref. \citenum{Masteran2023}, by addition of the HF CBS\cite{Froese1963} and the corresponding correlation CBS energy for CCSD\cite{Valeev2008}. 
      The CCSDT CBS energy was evaluated by addition of the difference between CCSDT-F12\cite{Shiozaki2009} and CCSD-F12\cite{Valeev2008} in the aug-cc-pV6Z basis to the CCSD CBS energy\cite{Valeev2008}.}
      \label{tab:ne_abs}
  \end{table}

Table \ref{tab:ne_contr} shows the different energy contributions of TC- and CT-F12-based CC methods.
First, we note that the mean-field contribution from the transcorrelated Hamiltonian is one magnitude larger than the CT-F12 contribution for the aug-cc-pVDZ basis set and is almost unaffected by the choice of the atomic orbital basis set, while the CT-F12 contribution approaches zero with an increasing number of basis functions.
The TC-CCSD-n energy contribution, while negative for the double-$\zeta$ basis, becomes positive for the core-valence double-$\zeta$ and triple-$\zeta$ basis sets, whereas the CT-F12-CCSD correlation energy contribution, as expected, decreases. 
Also, in contrast to the CT-F12-CCSDT results, the contribution from iterative triples is significantly smaller in the double-$\zeta$ basis for the transcorrelated approach and becomes positive for the larger basis sets. \\
Table \ref{tab:ne_abs} presents the resulting absolute energies for the different methods. 
Note that the difference between the energies does not match the entries in Table \ref{tab:ne_contr} due to the density fitting basis employed in this work. 
From the absolute energies, it becomes more evident that the transcorrelated methods approach the CBS from below. 
The application of the core-valence basis set, in combination with the TC-CCSDT-n ansatz yields the smallest error of 0.171318 hartree compared to the CBS, while CT-F12-CCSDT with the triple-$\zeta$ basis results in an error of only -0.020627 hartree. 
However, as discussed already for the other systems, the transcorrelated approach applied in this study works best with a core-valence basis, which is also apparent from the data for the Ne atom, where the error for the core-valence double-$\zeta$ basis set is even smaller as for the triple-$\zeta$ basis.

\subsection{LiH Potential Energy Curve}
We studied the LiH as a potential single-reference diatomic system. 
Hence, higher-order clusters should only have a small effect on the potential energy curve.
Like the Be atom, LiH is a four electron system, hence the CCSDTQ and FCI methods are equivalent.  

\begin{figure}[!htb]
    \centering
    \includegraphics[width=.9\linewidth]{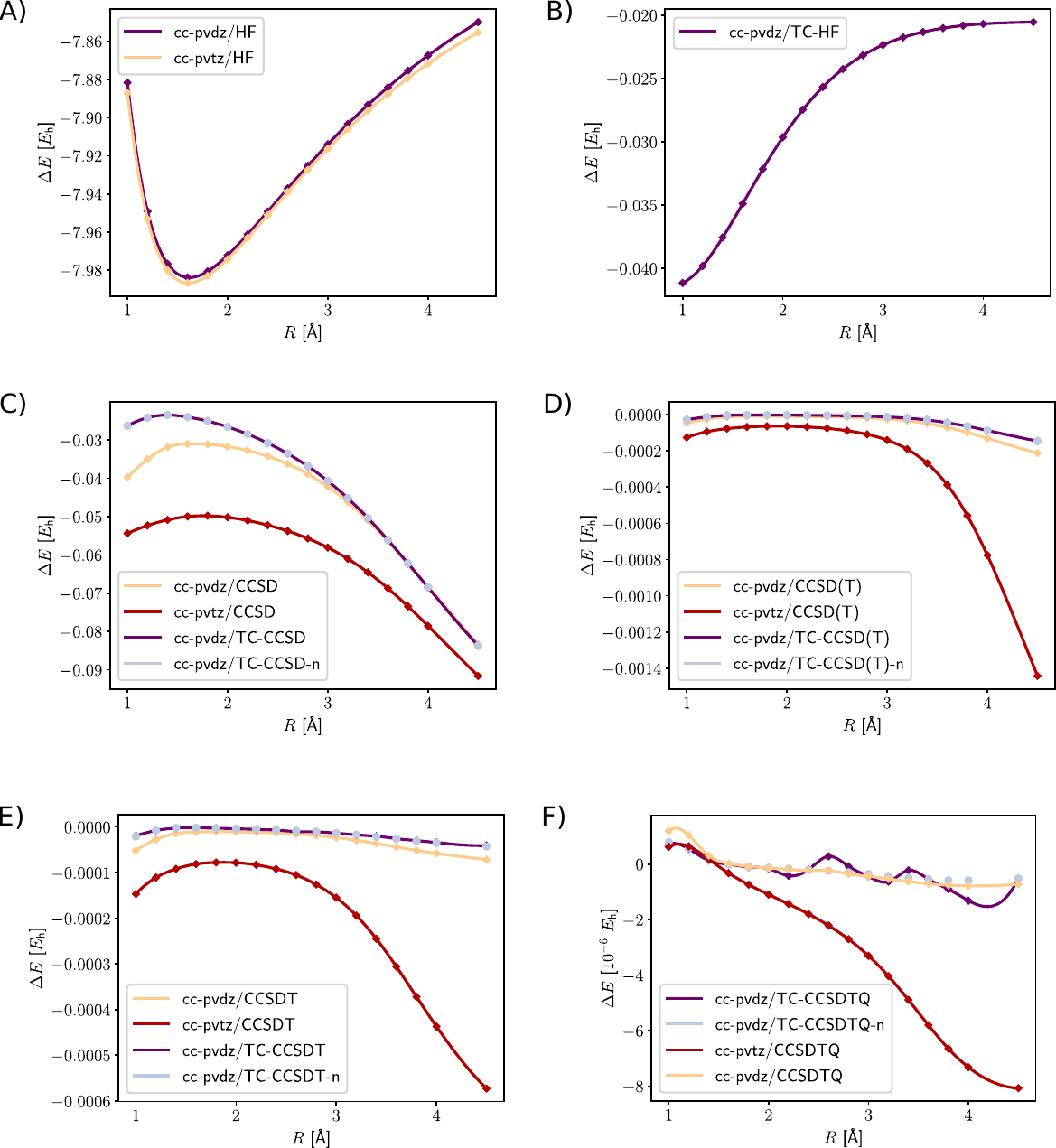}
    \caption{
    Dissociation of the LiH molecule:
    A) mean-field contribution of the conventional Hamiltonian, B) additional mean-field contribution of the TC Hamiltonian, C) CCSD correlation energy contribution, D) perturbative triples correlation energy contribution, E) full triples correlation energy contribution, F) full quadruples correlation energy contribution.
    The energy contributions in hartree ($E_\text{h}$) are plotted against the internuclear distance in \AA.}
    \label{fig:lih_dissociation}
\end{figure}

We compare results obtained with the transcorrelated with the conventional Hamiltonian for the different CC models in Fig. \ref{fig:lih_dissociation}. 
The pair correlation contribution from the transcorrelated Hamiltonian decreases for longer bond length, due to the electronic distance appearing in the correlator.
The correlation energy contribution of conventional CCSD is larger than for TC-CCSD around the minimum of the potential energy curve because the transcorrelated Hamiltonian already captures some of the correlation effects described by CCSD. 
However, for longer bond length the correlation energy contributions become nearly identical. 
Since the LiH dissociation can be reliably described by CCSD, higher-order cluster contributions, as in CCSD(T), CCSDT and CCSDTQ, do not improve the energy convergence significantly. 
Also the correlation energy contributions of the conventional and transcorrelated models show a similar trend in the cc-pVDZ basis.
As expected, the correlation energy contributions increase for the triple-$\zeta$ basis set.

The potential energy curves (see Supporting Information) confirm this, because the various CC methods resulted in a similar potential energy curve progression. 
We also observed similar trends as those found for the transcorrelated density matrix renormalization group in Ref. \citenum{szenes2024}.
First, the results for the transcorrelated Hamiltonian improves the basis set convergence by up to two cardinal numbers. 
Second, the effect of the normal-ordering approximation did not significantly affect the energy, which confirms the results from previous work\cite{Schraivogel_2021, szenes2024}. 
And third, the transcorrelated potential energy curves are not parallel to the conventional curves, which might be due to our choice of the correlation factor.
However, in contrast to other correlation factors, our choice is universal and simpler, which is an advantage for the definition of a universal electronic structure model. 
In view of our results with specific core-valence basis sets,
we presume that the parallelity issue can be solved by fitting proper atomic orbital basis sets for a transcorrelated approach, rather than introducing parameters into the correlators that may then be optimized (see also our results for the Be dimer in the next section).

Next we investigate the dissociation of the Be$_2$ dimer, to compare our perturbative triples approximation with respect to full triple excitations and to study the role of the basis set.

\subsection{Be$_2$ Potential Energy Curve}
A qualitatively correct description of the potential energy curve of the beryllium dimer with the conventional Hamiltonian requires at least a triple-$\zeta$ basis set and triple excitations in the CC expansion\cite{liu1980,harrison1983,magoulas2018,lesiuk2019}. 
To investigate the perturbative triples correction in combination with the transcorrelated Hamiltonian, we evaluated the Be$_2$ potential energy curve in the double-$\zeta$ and triple-$\zeta$ basis set. 
Furthermore, we compare results for the cc-pVDZ basis set with those obtained for the corresponding core-valence basis set cc-pCVDZ. 

\begin{figure}[!htb]
    \centering
    \includegraphics[width=.9\linewidth]{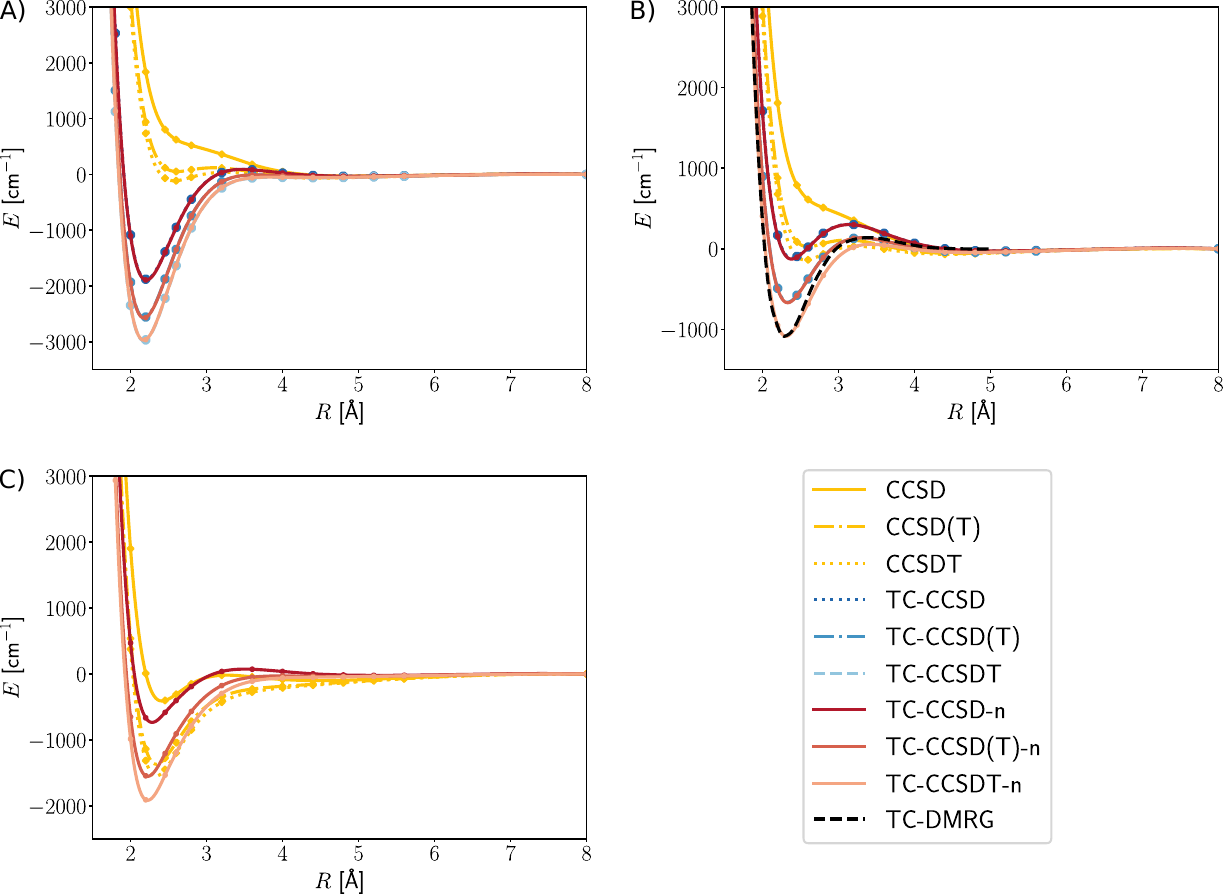}
    \caption{Dissociation of the Be$_2$ dimer in A) cc-pVDZ, B) cc-pCVDZ, and C) cc-pVTZ basis. 
    The energies are plotted in cm$^{-1}$ against the internuclear distance in \AA.
    The transcorrelated CC results are denoted with the prefix 'TC-', and the normal-ordering approximation is denoted with the suffix '-n'. 
    No prefix implies that the conventional Hamiltonian was used. 
    In B), we additionally compared with transcorrelated density matrix renormalization group (TC-DMRG) results from Ref. \citenum{Baiardi_2022}.}
    \label{fig:be2_dissociation}
\end{figure}

Fig. \ref{fig:be2_dissociation} shows that conventional CCSD results neither in the cc-pVDZ, nor in the cc-pCVDZ basis set yielded a qualitatively correct potential energy curve, because it is repulsive and no minimum could be identified.
The inclusion of perturbative and full triples resulted in a stable dimer, but produces an unphysical hump for internuclear distances slightly longer than the equilibrium bond length.
The same holds for conventional CCSD in a triple-$\zeta$ basis, whereas the inclusion of triple excitations improved the potential energy curve. 

The application of the transcorrelated Hamiltonian resulted in a well defined minimum already for the cc-pVDZ basis set in every CC model, but the curves are affected by a small unphysical hump. 
Transcorrelated CCSD(T) lowered the energy around the minimum and reduced the unphysical hump. 
The inclusion of full triple excitations further improved the simulation accuracy. 

As the basis set size increases, the difference between transcorrelated and conventional results decreases, since the transcorrelated Hamiltonian converges to the conventional Hamiltonian in the complete basis set limit.
Because of the size of the triple-$\zeta$ basis, we were unable to carry out full-transcorrelated Hamiltonian calculations in that basis. 
As expected, transcorrelated CCSD still delivered the unphysical hump, whereas transcorrelated CCSD(T) and CCSDT produced a qualitatively correct potential energy curve. 

Interestingly, in the core-valence double-$\zeta$ basis set, the bonding energy is even further reduced, even though the basis set size is smaller than the triple-$\zeta$ basis set. 
Accordingly, it has already been suggested to apply core-valence basis sets in the context of transcorrelation\cite{hino2001,Dobrautz_2022}.
The inclusion of nucleus-nucleus-electron correlation in the correlator may 
lift this requirement\cite{Cohen_2019}, which however, leads to a more complicated correlator. 
It therefore appears to be in order to optimize specific core-valence atomic-orbital basis sets for the application in transcorrelated calculations, to support a rather simple and universal correlation factor.

The transcorrelated CCSDT potential energy curve coincides with the transcorrelated density matrix renormalization group minimum from Ref. \citenum{Baiardi_2022}, but the unphysical hump is reduced for CCSDT, indicating that the bond dimension was too small. 

\begin{figure}[!htb]
    \centering
    \includegraphics[width=0.9\linewidth]{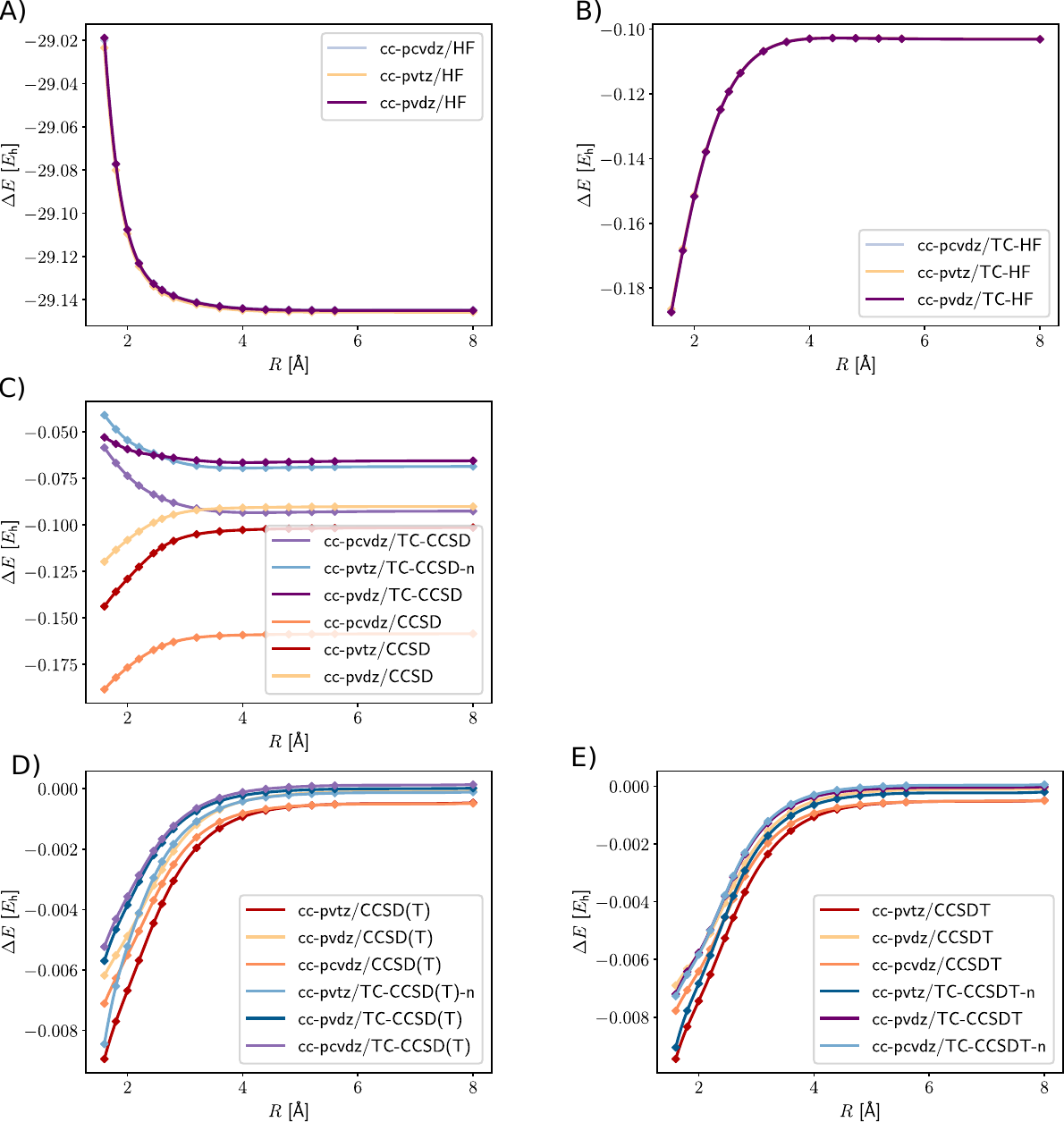}
    \caption{Dissociation of the Be$_2$ molecule: A) mean-field contribution of the conventional Hamiltonian, B) additional mean-field contribution of the TC Hamiltonian, C) CCSD correlation energy contribution, D) perturbative triples correlation energy contribution, E) full triples correlation energy contribution.
    If available, the energies, in hartree ($E_\text{h}$), of the full Hamiltonian are shown instead of the energies corresponding to the normal-ordered Hamiltonian, since they hardly differ from one another.}
    \label{fig:be2_contr}
\end{figure}

Plotting the individual contributions of the Be$_2$ dissociation in Fig. \ref{fig:be2_contr} highlights the dependence of the transcorrelated contribution on the interatomic distance.
While the correlation energy contribution from CCSD decreases with an increasing bond length for the conventional Hamiltonian, it increases for the transcorrelated Hamiltonian. 
This could be attributed to a counterbalance of transcorrelation, 
but the lack of the orbital optimization with the transcorrelated Hamiltonian must not be forgotten, because the transcorrelated integrals are then transformed with the conventional molecular orbital coefficients.
Additionally, while the contribution 
from transcorrelation changes only slightly for the different basis sets, the contribution from CCSD changes drastically. 
For the conventional Hamiltonian, the CCSD correlation energy contribution 
increases, as expected, with a higher cardinal number of the one-electron basis set
and the inclusion of core orbitals.
For the transcorrelated Hamiltonian, however, the energy contribution of CCSD 
for the double-$\zeta$ and triple-$\zeta$ is similar, because basis set effects are already included in the Hamiltonian and the transcorrelated mean-field contribution.
The contribution of the perturbative triples, as well as full triples, shows a similar trend for both, the conventional and the transcorrelated Hamiltonian.

\begin{figure}[!htb]
    \centering
    \includegraphics[width=.9\linewidth]{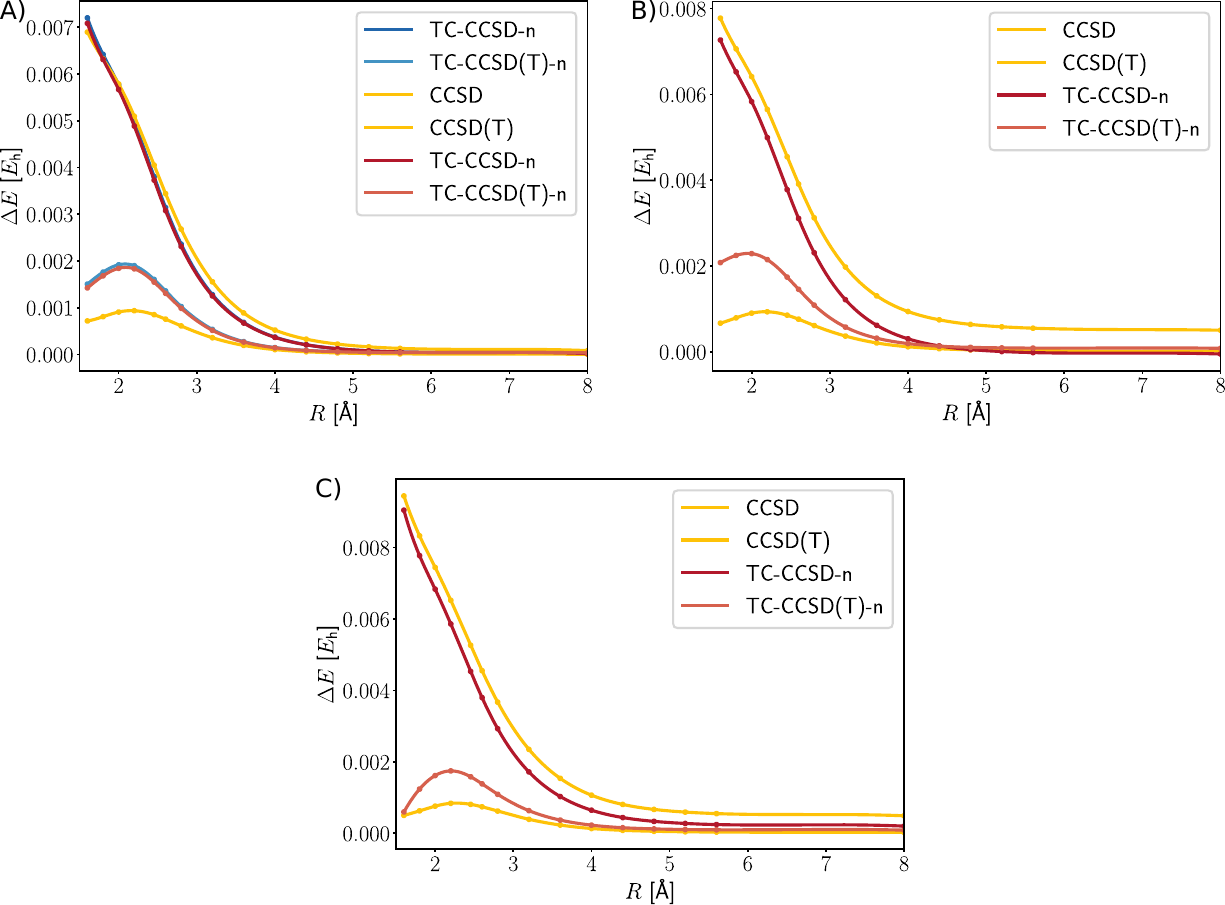}
    \caption{Difference of the correlation energy in hartree ($E_\text{h}$) between CCSDT and the other CC models along the dissociation coordinate in \AA{} of the Be$_2$ dimer.
    Basis sets: A) cc-pVDZ, B) cc-pCVDZ, and C) cc-pVTZ.
    The difference is evaluated for the same Hamiltonian, i.e. the curve for TC-CCSD-n shows the difference to the transcorrelated CCSDT energy with the normal-ordering approximation.}
    \label{fig:be2_dissociation_triples}
\end{figure}

Furthermore, we investigated the captured correlation energy of CCSD and CCSD(T) with respect to CCSDT (see Fig. \ref{fig:be2_dissociation_triples}). 

\begin{figure}[htb]
    \centering
    \includegraphics[width=.9\linewidth]{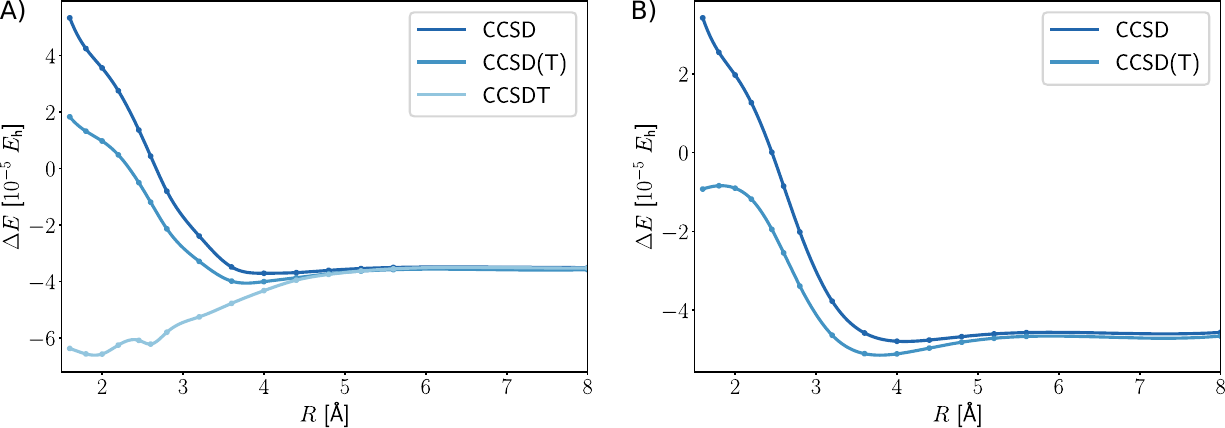}
    \caption{Error of the normal-ordering approximation for Be$_2$ dissociation in different CC models in A) cc-pVDZ and B) cc-pCVDZ basis sets.
    The energy difference is in hartree ($E_\text{h}$) and the internuclear distance in \AA. 
    }
    \label{fig:be2_dissociation_no}
\end{figure}

Fig. \ref{fig:be2_dissociation_triples} shows that the electron correlation of CCSD is captured equally for the conventional and transcorrelated Hamiltonian in the double-$\zeta$ basis. 
However, in contrast to conventional CCSD, the transcorrelated variant captured more correlation energy in the larger basis sets. 
By contrast, transcorrelated CCSD(T), however, recovered less correlation energy as the conventional counterpart, similar to the observations that we made for the Be atom. 
The potential reason is the lifted symmetry of the two-body operator in combination with the approximation of the left cluster operator as $\hat{T}^\dagger$.
Hence, a bivariational optimization, as described in Ref. \citenum{kats2024}, could potentially yield a lower error in the correlation energy.
Since the CC models presented in this study are all single reference, this also means that the main part of the electron correlation from the transcorrelated Hamiltonian is in the reference determinant. 
This also explains why the normal-ordering approximation works well, even in a dynamically correlated system such as Be$_2$ (see Fig. \ref{fig:be2_dissociation_no}). 

Generally the error of the normal-order approximation is negligible, with respect to the huge memory requirements of the three-body operator.

\section{Conclusions}
We derived a perturbative triples approximation for the transcorrelated Hamiltonian, including three-body couplings.
We showed that the required number of tensor contractions for CCSD, CCSD(T), CCSDT, and CCSDTQ for a Hamiltonian including a three-body operator drastically increases, in contrast to the conventional Hamiltonian.
Code generation tools are required for an implementation to tackle this challenge. 

Even for high-order CC the normal-ordering approximation, which 
introduces the mean-field contributions of the three-body functions in the lower-body operators and omits the remaining functions,
has a negligible effect on the energies.
Hence, the number of required tensor contractions and, therefore, the computational scaling can be reduced to that of conventional CC. 
Based on the Be atom and dimer, we can conclude that our transcorrelated CCSD(T) approximation improves transcorrelated CCSD, however to a smaller degree than for the conventional Hamiltonian.
The transcorrelated Hamiltonian roughly increases the basis set convergence by two cardinal numbers on the $\zeta$-contraction scheme, as we showed for the LiH dissociation and the fact that TC-CCSD(T) and TC-CCSDT have been able to produce a qualitatively correct potential energy curve for the dissociation of the Be$_2$ dimer.

We also showed that simple universal correlation factors, as the one applied in this work, can result in an unbalanced treatment of the electron correlation, 
which may be cured by fitting proper atomic orbital basis sets, in view of the results obtained for core-valence polarization basis sets.

A bivariational approach toward the amplitudes could even further increase the accuracy of the perturbative triples, because the de-excitation operator would match the amplitudes for the bra.

\section*{Acknowledgements}
M.R. and M.M. gratefully acknowledge support by ETH Research Grant (no. ETH-43 20-2). This work was presented at the Faraday Discussion "Correlated electronic structure" in London, UK, on 17 July.

\section*{Appendix}
As shown in Ref. \citenum{Baiardi_2022}, the following basic quantities are necessary to evaluate the three-body component of the transcorrelated Hamiltonian
\begin{align}\label{eq:three_body_ints_ex}
    K_{\mu\nu,\lambda \sigma,\kappa \tau} = 
    (\mu \nu | \nabla_1 f(r_{12}) |  \lambda \sigma | \nabla_1 f(r_{13}) | \kappa \tau)
\end{align}
where we have switched to the chemists’ notation for the sake of brevity with the positions of the sets $\mu\nu$, $\lambda\sigma$, and $\kappa\tau$ in the bra-ket correspond to electron coordinates 1, 2, and 3, respectively.
Here and throughout this section, Greek letters denote indices corresponding to the atomic-orbital basis. To make the evaluation of $K_{\mu\nu,\lambda \sigma,\kappa \tau}$ feasible,
it has been proposed~\cite{Baiardi_2022} to employ a combination of density-fitting (DF) and resolution of identity (RI) techniques. This leads to the following working formula
\begin{align}\label{eq:three_body_ints_app}
K_{\mu\nu,\lambda \sigma,\kappa \tau} \approx 
 C_{\mu \nu}^{P} C_{\lambda \sigma}^{Q} \alpha_{PQ}^{\kappa \tau} + 
C_{\mu \nu}^{P} C_{\kappa\tau}^{Q} \alpha_{PQ}^{\lambda\sigma} + 
C_{\lambda \sigma}^{P} C_{\kappa \tau}^{Q} \beta_{PQ}^{\mu\nu} - 
2 C_{\mu \nu}^{P} C_{\lambda \sigma}^{Q} C_{\kappa\tau}^{R} \gamma_{PQR}^{}
\end{align}
where the intermediates read
\begin{align}
    \alpha_{PQ}^{\kappa \tau} & = \left(P | \nabla_1 f(r_{12}) | Q | \nabla_1 f(r_{13}) | \kappa \tau \right) \\
    \beta_{PQ}^{\mu \nu} & = \left(\kappa \tau | \nabla_1 f(r_{12}) | P | \nabla_1 f(r_{13}) | Q \right) \\
    \gamma_{PQR}^{} & = \left(P | \nabla_1 f(r_{12}) | Q | \nabla_1 f(r_{13}) | R \right)
\end{align}
with the indices $P,Q,\ldots$ denoting elements of the auxiliary DF basis. 
Because the first operator acts on electrons 1 and 2 and the second operator on electrons 2 and 3, these intermediates obey the symmetry relations
\begin{align}
    \alpha_{PQ}^{\kappa \tau} & = \alpha_{PQ}^{\tau \kappa} \\
    \beta_{PQ}^{\mu \nu} = \beta_{QP}^{\mu \nu} & = \beta_{QP}^{\nu \mu} = \beta_{PQ}^{\nu \mu} \\
    \gamma_{PQR}^{} & = \gamma_{PRQ}^{}
\end{align}
After inserting the resolution of identity, the intermediates take the form
\begin{align}
    \alpha_{PQ}^{\kappa \tau} & = -\left(\nabla[Px] | f(r_{12}) | Q\right)
    \left(\nabla[\kappa \tau] | f(r_{12}) | x \right) \\
    \beta_{PQ}^{\kappa \tau} & = \left(\nabla[\kappa x] | f(r_{12}) | Q\right)
    \left(\nabla[\tau x] | f(r_{12}) | P \right) \\
    \gamma_{PQR}^{} & = -\left(\nabla[Px] | f(r_{12}) | Q\right)\left(\nabla R | f(r_{12}) | x \right)
\end{align}
where $x$ stands for the elements of the RI basis. As an unfortunate side effect of the RI approximation, the three-body integrals evaluated according to Eq.~\eqref{eq:three_body_ints_app} no longer strictly obey the same symmetry relations (with respect to exchange of the orbital indices) as the original integrals $K_{\mu\nu,\lambda \sigma,\kappa \tau}$. While this problem vanishes in the limit of a complete RI basis, it leads to technical complications in practice, such as increased storage requirements, for any finite RI basis. To fix this issue, we introduce the following symmetrized intermediates
\begin{align}
    \tilde{\beta}_{PQ}^{\kappa \tau} & = \frac{1}{4} \left(\beta_{PQ}^{\kappa \tau} + \beta_{QP}^{\kappa \tau} + \beta_{PQ}^{\tau \kappa} + \beta_{QP}^{\tau \kappa} \right)  \\
    \tilde{\gamma}_{PQR}^{} & = \frac{1}{2} \left(\gamma_{PQR}^{} + \gamma_{PRQ}^{} \right)
\end{align}
and use the redefined quantities $\tilde{\beta}_{PQ}^{\kappa \tau}$ and $\tilde{\gamma}_{PQR}^{}$ for the evaluation of Eq.~\eqref{eq:three_body_ints_app}. One can show by direct calculation that this approach leads to the correct symmetry of the approximate three-body integrals, i.e. identical to the original quantities $K_{\mu\nu,\lambda \sigma,\kappa \tau}$. 

%\bibliography{refs}
\providecommand{\latin}[1]{#1}
\makeatletter
\providecommand{\doi}
  {\begingroup\let\do\@makeother\dospecials
  \catcode`\{=1 \catcode`\}=2 \doi@aux}
\providecommand{\doi@aux}[1]{\endgroup\texttt{#1}}
\makeatother
\providecommand*\mcitethebibliography{\thebibliography}
\csname @ifundefined\endcsname{endmcitethebibliography}
  {\let\endmcitethebibliography\endthebibliography}{}

%\newpage
%\begin{center}
%\textbf{TOC Graphic}
%\begin{figure}[H]
%   \centering
%   \includegraphics[width=0.6\textwidth]{toc.pdf}
%   \label{TOC graphic}
%\end{figure}
%\end{center}

\end{document}